\PassOptionsToPackage{dvipsnames}{xcolor}
\documentclass[ijgi,article,submit,moreauthors,pdftex,10pt,a4paper]{mdpi} 

\firstpage{1} 
\articlenumber{x}
\doinum{10.3390/------}
\pubvolume{xx}
\pubyear{2017}
\copyrightyear{2017}
\history{Received: date; Accepted: date; Published: date}
\pdfoutput=1

\usepackage{graphicx}
\DeclareGraphicsExtensions{.pdf,.jpg,.png}

\usepackage{todonotes}
\usepackage{subcaption}
\usepackage{pdfcomment}

\Title{Historical collaborative geocoding}

\Author{R\'emi Cura $^{1,4,\dagger}$\orcidA{}, Bertrand Dumenieu $^{3,4,\dagger}$\orcidB{}, Nathalie Abadie $^2$, Benoit Costes $^2$\orcidD{}, Julien Perret $^{2,3,4}$\orcidE{}, Maurizio Gribaudi $^{3,4}$}
\AuthorNames{R\'emi Cura, Bertrand Duménieu, Nathalie Abadie, Julien Perret, Maurizio Gribaudi}

\address{%
	$^{1}$ \quad IGN; remi.cura@gmail.com\\
	$^{2}$ \quad IGN; first.last@ign.fr\\
	$^{3}$ \quad EHESS; last@ehess.fr\\
	$^{4}$ \quad GeoHistoricalData; first@geohistoricaldata.org}

\corres{Correspondence: julien.perret@ign.fr}

\firstnote{These authors contributed equally to this work.}  

\abstract{
	Latest developments in the field of digital humanities have increasingly enabled the construction of large data sets which can easily be accessed and used.
	These data sets often contain indirect spatial information, such as historical addresses.
	Historical geocoding is the process of transforming indirect spatial information into direct locations which can be placed on a map, thus allowing for spatial analysis and cross-referencing.
	There are many geocoders that work efficiently for current addresses. These do not, however, tackle  temporal information, and usually follow a strict hierarchy (country, city, street, house number, etc.) which is difficult, if not impossible, to use with historical data.
	Historical data is filled with uncertainty (pertaining to temporal, textual and positional accuracy, as well as to the reliability of historical sources) which can neither be ignored nor entirely resolved. 
	Our open source, open data, and extensible solution for geocoding is based on extracting a large number of simple gazetteers composed of geohistorical objects, from historical maps.
	Geocoding a historical address becomes the process of finding one or several geohistorical objects in the gazetteers which best match the historical address searched by the user.
	The matching criteria are customisable, weighted, and include several dimensions (fuzzy string, fuzzy temporal, level of detail, positional accuracy).
	Since our goal is to facilitate historical work, we also put forward web-based user interfaces which help geocode (one address or batch mode) and display results over current or historical maps. Geocoded results can then be checked and edited collaboratively (no source is modified).
	The system has been tested on the city of Paris, France, for the 19\textsuperscript{th} and 20\textsuperscript{th} centuries.
	It shows high response rates and works fast enough to be used in an interactive way.
}


\begin{document}
	
	%
	
	\section{Introduction}
	\subsection{Context}
	
	In historical sciences, cartography and spatial analysis are extensively used to uncover the spatial patterns at play within textual historical data. This data contains indirect textual references about location, such as place names (toponyms) or postal addresses.
	In order to map such data, each item needs to be geocoded, \emph{i.e.} assigned with coordinates through the matching of an indirect spatial reference with entities identified in a geographical data source (\emph{e.g.} a map georeferenced in a well-known coordinate reference system)~\cite{Goldberg2007}.
	Problems emerge when such spatial references become obsolete due to the temporal gap between the data to be geocoded and the reference datasource: locating the London Crystal Palace (destroyed by fire in 1936) on a current map would be rather tricky.
	Worse still, it might create ambiguities and possibly lead to erroneous geocoding, as the Crystal Palace nowadays refers to a South London residential area.
	Although manual geocoding can help in such cases, the constantly increasing volume of historical data, which results from the flourishing number of initiatives in the field of digital humanities, calls for automatic approaches.
	Despite the existence of highly efficient geocoding tools and API for modern data, it remains a challenge to come up with a truly historical geocoder.~\cite{St-Hilaire2007,daras2014hag}.

	\subsection{Approach and contributions}
	
	The main focus of this article is the historical geocoding problem: providing the best matching geohistorical objects in available gazetteers for a given textual address query.
	
	We propose to depart from the classic geocoding paradigm, where a high quality, hierarchical, complex and complete gazetteer is used in conjunction with a simple matching method.
	Instead, we intend to relax the constraints on the address definition process, and use several simpler gazetteers at the same time. The complexity is transferred to the matching method, which is fully temporal, fuzzy, and can be customized according to the user's goal.
	
	We also discuss the construction of a geohistorical database, the development of data matching (linkage) methods which make full use of the temporal aspects of geohistorical data and the input query, as well as the collaborative dimension. 
	The main contributions of this article consist in
	(1) a formalisation of the historical geocoding problem,
	(2) a minimal model of geohistorical objects which can easily be re-used and extended,
	(3) an open source geocoding tool which is powerful, easy to use and can be extended with any geohistorical data,
	(4) a graphic tool to control and edit the geocoding results, which can then optionally be used to enrich the geohistorical database,
	(5) a qualification of geocoding results in textual, spatial, and temporal terms.

	\section{Theory}
	
	\subsection{Geocoding}
	
	\subsubsection{Related work}
	Geocoding is an inevitable step in any spatially-based study with considerable bodies of data. This makes it a critical process in various contexts: public health, catastrophe risk management, marketing, social sciences, etc.
	Many geocoding web services have been developed to fulfil this need, originating from private initiatives (Google Geocoding API, Mapzen\footnote{\url{mapzen.com}}), public agencies (the French National Address Gazetteer\footnote{\url{adresse.data.gouv.fr/api/}}) or from the open-source community (OSM Nominatim\footnote{\url{nominatim.openstreetmap.org}}, Gisgraphy\footnote{\url{gisgraphy.com}}).
	These services can be characterized in terms of their three main components~\cite{Goldberg2007,Hutchinson2013}: input/output data, reference dataset and processing algorithm.
	The \emph{input} is the textual description the user ambitions to refine into coordinates. It might take the form of a traditional address containing a building number, street name, city name, or country (\emph{e.g.} "13 rue du Temple, Paris, France"), but it may also be incomplete, or simply refer to a landmark (\emph{e.g.} "The Eiffel Tower, Paris").
	The~\emph{reference dataset} designates a gazetteer which pairs names of geographical entities (places, addresses) with geographical features. 
	Because the main geocoding tools are provided by heavyweight actors of geographical information such as Google, Microsoft, OSM or the national cartographic agencies, the geographical databases they produce are used as the reference dataset for these geocoders.
	These databases are extremely structured (hierarchy, normalization) and of high quality. 
	
	The \emph{processing algorithm} consists in finding the best-matching element from the reference dataset for the associated input description.
	Finally, the output usually contains a geographical feature along with its similarity score (e.g. perfect or approximate match).
	Although the geometries of the matched features may be complex, they are most often rendered into simple two-dimensional points.
	
	\subsubsection{Estimating and conveying of the quality of geocoded places}   
Because of its ability to transform the indirect spatial reference of a piece of information into a direct spatial reference, the process of geocoding is a critical stage of many spatial analyses wherein data is not directly associated with geographical coordinates (e.g observations associated with place names).
However, geocoding cannot be limited to this process: it is crucial to estimate or measure the quality of each individual indirect-to-direct transformation and either convey this information along with the final results or provide a mechanism which can correct these results.
It would otherwise be impossible to establish a distinction between result variations due to the imperfection of the geocoding process and a real phenomenon which could be hidden in the data.

    The quality of geocoding services can be estimated via two very important criteria~\cite{Roongpiboonsopit2010}.
	First, the database quality: how complete and up to date is the reference database?
	Second, result characterisation: how spatially accurate is each result and what is its associated reliability?
	In addition, the quality of the matching process can also be evaluated (how the process deals with errors in the input address, for instance).
	
	\subsubsection{Temporal depth} 
	Common geocoding approaches cannot be used for (geo)historical data for three main reasons.
	To begin with, existing geocoding services do not take the temporal aspect of the query, or the dataset they rely on, into account.
	Indeed, they usually rely on current data, such as \emph{OpenStreetMap}\footnote{\url{openstreetmap.org}} data, which is continuously updated.
	As such, they implicitly work on a valid time that is the present (or possibly the interval between the beginning of the database construction and present time).
	The second reason is that they rely on an exhaustive, strongly hierarchical database whose accuracy can be checked against ground truth (\emph{i.e.} there is always a way to check the actual location of an address, the database can therefore constitute an unambiguous and objective reference). 
	Unfortunately, historical data cannot easily be verified: one has to compare it with different available (geo)historical sources (possibly incomplete and conflicting) and must often make assumptions or hypotheses. Such hypotheses are in turn continuously challenged and updated by new discoveries, and there is no way to provide a truly definitive answer.
	Primary sources may also be wrong or misleading. Modern geocoding tools are not geared towards dealing with these ambiguities. 
	Finally, available historical sources for the weaving of a gazetteer are sparse (both spatially and temporally), heterogeneous, and complex.
	We believe all these specificities call for a dedicated approach.
	Similar observations have already been made in the context of archival data by the UK National Archives, for example~\cite{Clough2011}. 
	Large historical event gazetteers already exist~\cite{Mostern2008, southall2011} and provide an important basis to the development of the reference dataset.
	More specifically, the classic steps we have identified in geocoding for a historical source are first to establish a reference gazetteer for addresses (associating standardized textual addresses with coordinates), and then to determine the input standardized textual addresses within this gazetteer (geocoding).
	Theoretically speaking, this whole process is very akin to a simple database join where the key would be the standardized textual address.
	However, this methodology does not take into account the historical dimension, and requires both standardized and complete gazetteers. 
	For instance, \citep{St-Hilaire2007} pinpoint a reference gazetteer of "CSD" (census subdivision) for each year in the historical period of interest. Geocoding is then completed separately for each year, with a simple CSD match.
	\citep{logan2011} introduce more detailed address gazetteers (which go so far as to provide street numbers). However, since the work was still in progress at the time the article was written, it gives no details on the geocoding process.
	\citep{lafreniere2015} present work pertaining to several historical periods, and for each one, comprises a gazetteer of standardized addresses. When an historical address needs to be geocoded, the temporally closest gazetteer is chosen, which then allows for regular matching to occur.
	
	Our ambition, however, is to fully make use the time dimension, and to relax the constraints on gazetteers.
	First, we can simultaneously combine several simple gazetteers by merely using a minimal subset of required information (according to the suggested geohistorical object model). Each gazetteer contains data which can both be fuzzy (errors in the address text, the date, or the position) or situated on different scales (house address, street, neighbourhood). Moreover, gazetteers may conflict with one another.
	Then, the geocoding of a query address is achieved thanks to a sophisticated multi-dimensional matching tool which can be customized according to user needs.
	
	\subsubsection{Handling the imperfections of geohistorical data} 
	Geohistorical data, as any other type of data, contains imperfections.
	Such imperfections can be categorized into 3 main classes~: uncertainty, imprecision and ambiguity~\cite{de2008imperfection}.
	Uncertainty applies to information of which the reliability can be questioned:
	To what extent can we trust the location of an address point depicted in a map, when we know that this map contains errors?
	Just as a GPS can generate a location within a 20 meter radius, the precision of the locations spawned within a historical map is limited by the precision of the map itself.
	Ambiguity arises from two situations. 
	First, different sources can provide conflicting information about the same geographic entity.
	Second, the information available in an entity can be too sparse to properly define the properties of that entity and can therefore be unable to produce data of sufficient quality.
	
	To our knowledge, no other historical geocoding approach has taken the the characterisation of geocoding results into consideration.
	However, it is an essential aspect for historical geocoding due to the very unprecise and sparse nature of geohistorical data.
	Indeed, geocoding results need to be validated and/or edited manually.
	
	Given the large amount of addresses (more than $100,000$ addresses for Paris) and the potential complexity of the task, this is clearly a lot of work.
	Fortunately, several projects such as \emph{OpenStreetMap} have lead the way for what is usually called \emph{Volonteered Geographical Information} (VGI)~\cite{Goodchild2007} of \emph{crowdsourcing geospatial data}~\cite{Heipke2010}.
	This approach consists in using collaboration to solve a problem collectively, usually by having citizens participate in the process.
	Such an approach has already extensively been used for historical data, although in distinctively different contexts.
	For instance, \citep{southall2017} put forward a website which aspires to collaboratively input the placenames that appear on the map of Great Britain for the years 1888 to 1914.
	Other projects such as Keweenaw history\footnote{\url{http://www.keweenawhistory.com/}} and several projects heralded by the New York City Public Library labs \citep{vershbow2013} have been using crowdsourcing to create or edit historical data, such as building footprints.
	Our approach is similar: a convenient web interface and the power of collaborative editing are also used. However, our end goal is different. Our purpose is not to create an authoritative historical data source, rather, we intend to allow each user to adapt the source to his or her own usage.

	As suggested in a recent typology of participation in citizen science and VGI~\cite{Haklay2013}, different levels of participation can be defined.
	These levels go from ``crowdsourcing'', where the cognitive demand is minimal, to ``extreme citizen science'' or ``collaborative science'', where citizens are involved in all the stages of a research (problem definition, data collection and analysis).
	In the rest of this article, we propose a collaborative historical geocoding approach for a simpler participation of citizens in geohistorical research using dedicated interactive tools.
	A reproducible research approach using open source tools and open data~\cite{Fomel2009, Aruliah2012, Wilson2016, Marwick2016} leads to a more collaborative historical science.

	\subsubsection{Handling heterogeneous types of addresses}
	Some modern geocoders are able to return various types of geographic features. 
	For instance, the Open Street Map geocoder can return a set of hierarchically organized geographical features.
	
	Similarly, the method we present here is able to return different geographical features based on the best match for the textual address.
	For example, the following textual address, "12 rue du Temple, Paris, France", might return a dot representing the building, a polyline representing the street, a polygon for the city or the country, depending on the available information and on user preferences (the scale parameter).

	\subsection{Integrating geohistorical data}
	
	
	\subsubsection{General considerations about building a spatio-temporal database}
	Extracting information from historical maps amounts to building a spatio-temporal database.
	There are several approaches to do so, and we stress that our attempt is not to create a continuous spatio-temporal database.
	Instead, we store representations of the same space at multiple moments in history, according to the well-known snapshot model~\citep{armstrong1988}.
	
	\subsection{Extracting geohistorical objects from historical maps}
	The starting point for building gazetteers is to extract information from historical maps.
	The first part of the extraction process is to scan the maps (i.e. going from a paper map to a computer file) and to georeference the map in a defined coordinate reference system.
	These maps are historical sources, and, as such, a historical analysis is performed in order to estimate the probable valid time (temporalisation), positional accuracy, completeness, confidence, relation to other historical maps, etc). 
	In our approach, we focus on geometrically accurate historical maps as our primary source for two main reasons:
	\begin{itemize}
		\item Historical maps are spatially close to modern maps. The way spatial information is described is very similar (both are based on mathematically well-defined reference systems, as opposed for instance to an artistic painting of a city which would be seen as a non-geometrical map). The integration of the information they convey in a Geographical Information System (GIS) is therefore facilitated.  
		\item The main goal of such maps is to provide a reliable depiction of the shape and location of geographical features.
	\end{itemize}
	Although this choice seriously reduces the number of possible sources and therefore lessens the quantity of accessible spatial information, it aims at efficiency.
	Indeed, geometrical maps are a good compromise because they are reliable while at the same containing a lot of spatial information, and can bear the complexity of information extraction.

	\subsubsection{Georeferencing historical maps}
	We must establish a correspondence between each pixel of the historical map and its geographical coordinates.
	To do so, we first choose a common spatial reference system (SRS).
	We then identify common geographical features between historical maps and current maps: so-called ground control points (GCP).
	Last, we compute a warping transform which will stick as closely as possible to the matching points.
	Finding GCPs between current maps and historical maps can be increasingly difficult as we go back in time, because there are less and less perennial GCPs.
	Consider, for instance, the city of Paris, where the French Revolution and its consequences combined with 19\textsuperscript{th} century transformations (including the so-called Haussmannian transformations) resulted in massive changes in the shape of the city.
	To this end, we can start by georeferencing \emph{e.g.} 20\textsuperscript{th} century maps to current maps, then georeference \emph{e.g.} 19\textsuperscript{th} century maps to 20\textsuperscript{th} century maps, and keep going for even older maps.
	A more in-depth analysis of the spatial quality of historical maps of Paris can be found in~\citep{Dumenieu2018}.
	
	\paragraph{\emph{Choosing the target spatial reference system}}
	Geographic coordinates are expressed through a coordinate reference system, which can either be geographical (i.e. coordinates are latitude and longitudes) or projected on a plane.
	Georeferencing a map requires choosing a target coordinate system to place it on the Earth's surface.
	It can be chosen arbitrarily, but it is advisable to select a coordinate system associated with a cartographic projection close to that of the map which is to be georeferenced.
	Indeed, most western countries' maps since the 18\textsuperscript{th} century have attempted to depict a geometrically accurate geographical space, which implies using a mathematical model for the Earth and to display a projection on a piece of paper. 
	Large-scale maps such as city maps usually rely on a simple Plate-Carr\'e projection with an approximation of the Earth, depicted as a flat surface.
	In the case of low-scale maps such as country maps, the projection and coordinate system depends on the state of geodetic knowledge and cartographic methods.
	In most cases, however, the exact parameters of the historical map projection are unknown.
	Ignoring the original projection and coordinate system of the map can result in geometrical distortions of the georeferenced map.
	
	\paragraph{\emph{Selection of ground control points}}
	The identification of pairs of GCPs is a critical step because the number, distribution and quality (\emph{i.e.} positional accuracy, reliability, confidence) of the points strongly influence the quality of the georeferencing.
	The reliability of the chosen GCPs actually depends on the geographic entities they are placed on, which calls for an in depth study of the construction process of the historical map.
	Because this can be a very time-consuming task, it is possible to choose the GCPs based on some simple rules.
	First, the GCPs should be located on the geographic entities that are the most stable through time.
	This typically includes the main religious and administrative buildings such as churches and palaces.
	On high-scale maps of cities, street intersections might also be acceptable supports for GCPs.
	On low-scale geometric maps, bell towers are often the most accurate objects since they have been extensively used as anchors for survey operations.
	In general, unstable geographical features such as rivers, forests, rural roads, coastal lines, etc. should be avoided.
	While the quality of the selected points depends on each map, a simple rule of thumb is to select as many homogeneously distributed points as possible in order to make some progress ~\cite{herrault2013}.
	Three parameters have to be considered: the geometrical type of the features carrying the ground control points, their nature and the method used to identify them.
	Usually, features chosen as ground control points are represented by 2D points; lines or surfaces may also be used, and possibly even curves~\cite{Fabbri2010}.
	For historical maps, the positional accuracy of mapping themes can greatly vary, either because of the map's purpose, or due to the mapmaking process itself.
	Optionally, geodetic features drawn on the map such as meridians or parallels can also be used as GCP, provided their geodetic characteristics can be fully specified (e.g. identify exactly which meridian is drawn in which exact reference system).        
	The actual identification of GCPs can be achieved by automatic or manual processes.
	Automatic approaches are notably used for historical aerial photographs, where feature detection and matching algorithms are well fitted~\citep{cleri2014}.
	Common GIS tools offer georeferencing software allowing to manually select pairs of \emph{ground control points} identified in both the input and the reference maps.
	Such tools are often used for historical maps georeferencing because: (1) they are easy to utilize and (2) they allow historians to control the quality and reliability of the identified points by using co-visualization between both maps. The NYPL Lab even proposed a web-based version geared toward historical data (Map Warper \footnote{\url{ http://maps.nypl.org/warper/}}).
	\paragraph{\emph{Choosing a geometric transformation model}}
	Once an acceptable set of paired features has been identified, the last step is to compute the transformation from the input map to the reference.
	Several transformation models have been proposed: global transforms (affine, projective), global with local adaptations (polynomial-based) and local transforms (rubbersheeting, Thin Plate Spline, kernel-based approaches, etc.).
	Studies have been conducted to assess the relevance of these transformations for historical maps~\cite{bitelli2009,boutoura2006,herrault2013}.
	They show that choosing a model is mostly a matter of compromise between the final spatial matching between the feature pairs (\emph{i.e.} the expected residual error) and the acceptable map distortion with regards to its legibility.
	Exact or near-perfect matching between features can be achieved with local transforms and high order polynomials, whereas the internal structure of the map is mostly preserved by global transformations.
	Low order polynomials offer a compromise between both constraints.
	
	\subsubsection{Temporalization: locating geohistorical sources in time}
	Georeferencing is a way of locating multiple maps in the same reference space.
	Similarly, \emph{temporalization} is the process of locating each geohistorical source in time.
	When building spatio-temporal snapshots from historical maps, the key problem is to determine the moment where the map is representative of the actual state of the area it portrays, \emph{i.e.} the \emph{valid time} of the map.
	We define the valid time of each map as the period starting with the beginning of the topographic survey and ending with the publication of the map, which is often uncertain.
	Representing uncertain or imprecise periods of time is a common issue when dealing with historical information and many authors relied on the fuzzy set theory to represent and reason on imperfect temporal knowledge~\citep{derunz2010,kauppinen2010}. In all generality, temporal knowledge is represented by a function of time with values ranging from 0 (the source provides no information at this time) to 1 (geographical entities portrayed in the map are regarded as existing and tangible at this time).
	
	\subsubsection{Extracting information from maps}
\begin{figure} [htb!]
		\begin{center}
			\includegraphics[width=\linewidth,keepaspectratio]{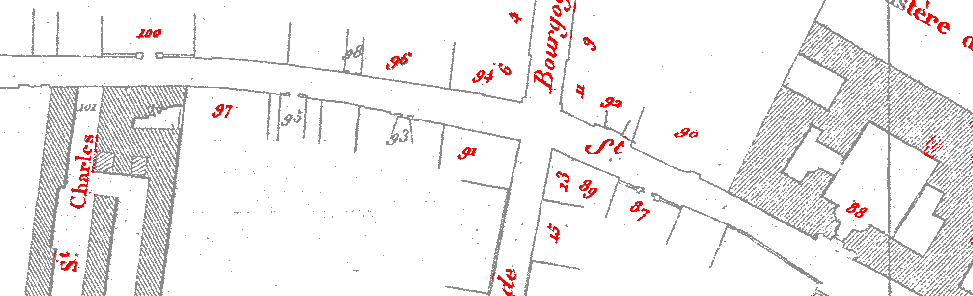}
			\caption{In this example, handwritten text is automatically detected and extracted (red) from a historical map using various image processing methods. Note that some building numbers are not extracted.}  
			\label{method_hsource_auto_extract}
		\end{center}
	\end{figure}

	Once the historical maps have been georeferenced and temporalized, their cartographic objects can be extracted to produce geohistorical objects.
	The most common way to extract information from maps is by human action with a classic GIS software (\emph{e.g.} QGIS).
	However, each historical map of Paris contains a large amount of information to be extracted (\emph{e.g.} thousands of street names, even more building numbers, etc.).
	A first solution would therefore be to use computer vision and machine learning methods to create automatic extraction tools.
	These tools can process the whole map in a few hours.
	Regrettably, such tools are difficult to design, are very specific to each historical map, and may produce low quality results (see Figure~\ref{method_hsource_auto_extract}).
	Collaborative approaches have recently shown to be very efficient for building large geographical databases in a relatively short period of time (OSM\footnote{\url{http://www.openstreetmap.org}}, NYPL\footnote{\url{http://buildinginspector.nypl.org/}}).
	In the end, for the use case of Paris, data is mainly extracted manually by experts, except for the Open Street Map data which is a mix of collaborative editing and collaboration with the French Mapping Agency (IGN).
	
	\section{Methods}
	
\begin{figure}[htb!]
		\begin{center}
			\includegraphics[width=\linewidth,keepaspectratio]{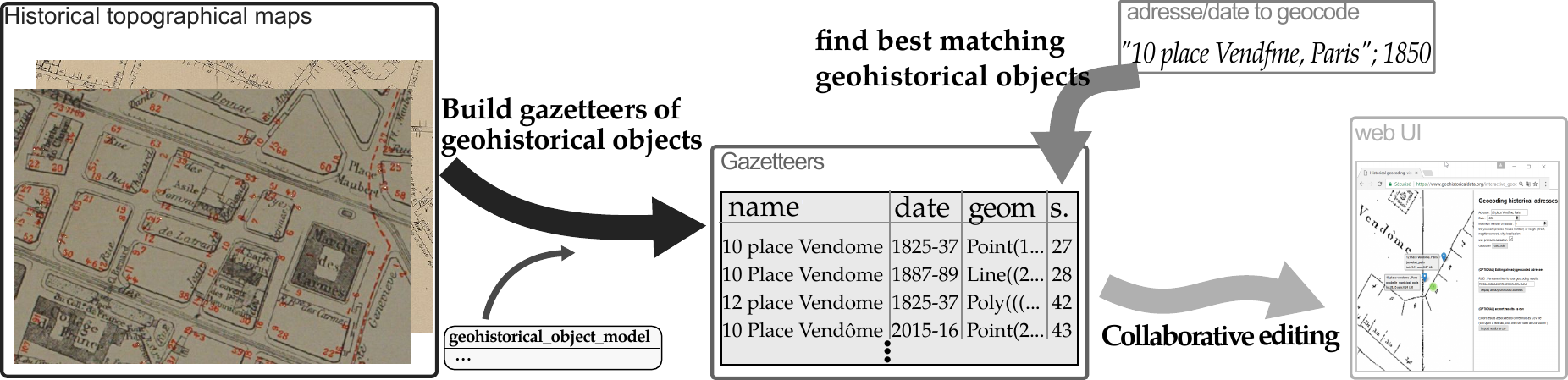}
			\caption{
				Gazetteers of geohistorical objects are created based on information extracted from georeferenced historical maps.
				Geocoding a historical address means finding the best matching object in these gazetteers, based on a customised function (semantic, temporal aspect, spatial precision, etc.).
				Results can be displayed through a dedicated web interface for collaborative editing.
			}
			\label{graphical_abstract}
		\end{center}
	\end{figure}

	Based on historical sources and historical maps, we extract geographical features which are then gathered into several gazetteers.
	These (geo) historical features are modelled in a generic way (geohistorical objects) into a Relational Database Management System (RDBMS). 
	Geocoding an input historical address is finding the geohistorical object in the gazetteers that best matches this historical address. We propose a matching process relying on several distances (temporal, textual, spatial, etc.) which can be customised by the user.
	Lastly the results can be displayed via a web mapping interface over current or historical maps, and further checked and edited collaboratively. Any edit creates a duplicate of the original geohistorical object from the gazetteer, which is then added as another geohistorical source (user contributed). 
	Figure~\ref{graphical_abstract} illustrates this approach.

	\subsection{Building historical gazetteers}
	Our approach for building a historical gazetteer follows these steps:
	\begin{enumerate}
		\item a historical map is scanned,
		\item scans are georeferenced using hand picked control points,
		\item historical work allows for the estimation of temporal information and spatial precision of the map,
		\item road names and axis geometry are extracted from the scan (manually or automatically),
		\item building numbers are extracted from the scan (manually or automatically),
		\item in some cases, building numbers can be generated from the available data (e.g. road starting and ending building number),
		\item normalised names are created from historical names (dealing with abbreviations, etc.),
		\item geohistorical objects are created.
	\end{enumerate}

	\subsubsection{Extracting geo-historical information from maps}
	The whole process is carefully designed and explained in detail in~\cite{Dumenieu2015} (A work on modelling historical geospatial information, from the source qualification to georeferencing to analysis and data extraction, associated with optimization methods to create and exploit spatio-temporal street graphs (linkage between historical information).
	
	Geohistorical objects are then extracted from the referenced historical maps, mostly manually (in a collaborative way), or with the help of computer vision techniques. 
	The main advantage is that for a given moment in time we can have several conflicting snapshots that coexist.
	This is essential, as solving these conflicts may not be possible, and reporting these several conflicting geocoding results to historians may help appreciate the results.
	The drawbacks of this model, \emph{i.e.} information redundancy and the inability to store the changes themselves, can be overcome during the geocoding process.
	
	\subsection{Modelling geohistorical objects}
	Information extracted from historical maps is used to create gazetteers. Gazetteers may contain different kinds of information, however we design a core set of information that these gazetteers have to possess: the geohistorical object model.
	To this end, we design a geohistorical object model with all the necessary attributes and the flexibility to adapt to a great variety of geohistorical object types and sources.
	Our goal is to provide a generic minimal (geo)historical object model which can be used by others and easily extended when necessary.
	Please note that this geohistorical object model is separate from the geocoding issue, and that several gazetteers may contain redundant or conflicting geo-historical objects.
	Such occupancy is allowed as it is common for historical sources to be redundant and conflicting. 
	Furthermore, the geocoding method is designed to take these issues into account.
	
	\subsubsection{Modelling geohistorical objects}
	Geohistorical data is extremely diverse, both in terms of historical sources and of how the sources were dealt with by historians.
	As such, historians use complex tailored models.
	We do not aim at modelling every geohistorical data in its own specificity and complexity.
	Instead, we propose to model the bare minimal common properties of all geohistorical objects, and offer mechanisms in order for this model to be easily extended and tailored to the specificities of the data. 
	To define the bare minimal model, we start from the very nature of a geohistorical object: both a historical object and a geospatial object.
	The extension mechanism is provided via a database-object oriented design using table inheritance, and is packaged into a PostgresSQL extension\footnote{\url{https://github.com/GeoHistoricalData/geohistorical_objects}}.
	
\begin{figure} [htb!]
		\begin{center}
			\includegraphics[width=\linewidth,keepaspectratio]{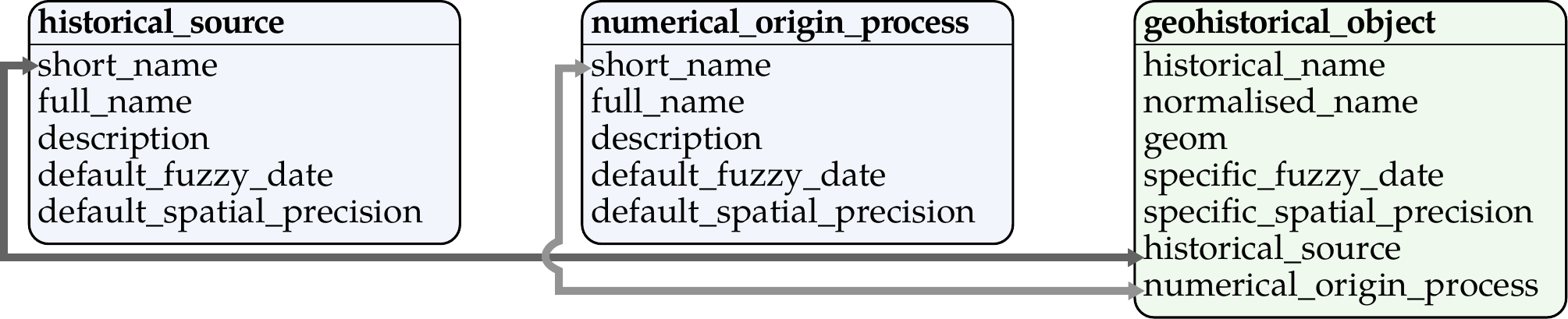}
			\caption{The geohistorical object model, where each object is characterized by its historical source (for instance the historical map the object was described in) and a numerical origin process, which is the process through which the object was digitized. Aside from source and origin processes, an object is also described by a fuzzy date, a text and a geometry.}  
			\label{geocoding_method_geospatial_obeject_model}
		\end{center}
	\end{figure}

	Geohistorical objects possess both a historical and a geospatial component.
	We stress that modelling the primary source and the extraction process of a geohistorical object is important in order to trace the provenance of the information.
	The details of the model are illustrated in Figure~\ref{geocoding_method_geospatial_obeject_model}.
	
	\paragraph{Historical aspect}
	In our model, a historical object is defined by its name, source and temporalization.
	\begin{itemize}
		\item \emph{Name.}
		By name, we mean the historical name initially used to identify the object in the historical source, and the current name used by historians to identify the object in the current context.
		For instance, the historical name for the Eiffel Tower in Paris may be "tour de 300 mètres", but today, it is referenced as "Tour Eiffel".
		Both can coexist in a gazetteer (two different geohistorical objects, with a different source and date).
		\item \emph{Source.}
		A historical object is defined by a primary historical source (document), where the object is referenced.
		Beside the historical source, the way the object was digitized in this source is also essential.
		For instance, a street name may have the Jacoubet map as its historical source, and would have been digitized via collaborative editing on the georeferenced map.
		\item \emph{Temporalisation.}
		Any historical source is associated with temporal information (fuzzy dates), which is the period during which the source is most likely to be relevant.
		Beside the historical source's temporal information, a historical object can also have its own temporal information.
		For instance, a street may have been extracted from a historical map created between 1820 and 1842.
		Using other historical documents may allow to narrow the probable existence of this street to 1824-1836. Keep in mind that several other geohistorical objects may describe this street at several other time periods in the same or in another gazetteer.
	\end{itemize}
	
	\paragraph{Geospatial aspect}
	A geohistorical object is also defined by geospatial information: a direct spatial reference (geometry) and its positional accuracy metadata.
	\begin{itemize}
		\item \emph{Geometry.}
		A feature has a geometry which follows the OGC standard\footnote{\url{http://www.opengeospatial.org/standards}}.
		It may be a point, polyline, polygon, or a composition of any of these, in a specified SRS.
		The geometry is extracted from the historical source in a manual or automatic way. Such information will be given in the \emph{Source} description.
		\item \emph{Positional accuracy.}
		Historical features have positional accuracy information.
		This precision expresses the spatial uncertainty of the historical source (the person drawing the map might have made mistakes) and the spatial imprecision of the digitizing process (the person editing the digitised map might have made a mistake).
		One historical source may contain several accuracy metadata, one for each geohistorical object type it contains.
		For instance, a historical map may contain buildings and roads.
		Buildings may have a different positional accuracy (5 metres) than road axis (20 metres).
		Besides, the digitising process precision may have been of 5 metres. 
	\end{itemize}
	
\begin{figure}[htb!]
		\caption{An uncertain valid time modelled as a trapezoidal fuzzy set function}
		\begin{center}
			\includegraphics[width=0.7\linewidth,keepaspectratio]{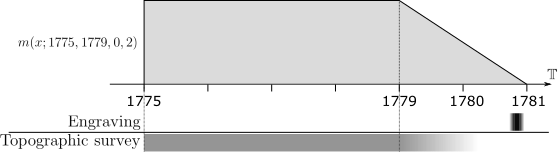}
		\end{center}
		\label{figure:fuzzytime}
	\end{figure}

	\paragraph{Temporal aspect}
	A historical source contains information about its valid time. This valid time is represented in a fuzzy way.
	Our model can adapt to any piece-wise linear function, but we chose to model imprecise valid times as trapezoidal fuzzy sets, since these functions are simple to understand, use, and cover most common use cases. 
	We rely on the pgSFTI\footnote{\url{https://github.com/OnroerendErfgoed/pgSFTI}} postgres extension to store and manipulate such temporal fuzzy information.
	For instance, Figure~\ref{figure:fuzzytime} illustrate the \emph{valid time} of a map whose topographic survey started in year 1775, ended between years 1779 and 1780 and which was engraved in late 1780.
	
	\subsubsection{A database of geohistorical objects}
\begin{figure} [htb!]
		\begin{center}
			\includegraphics[width=\linewidth,keepaspectratio]{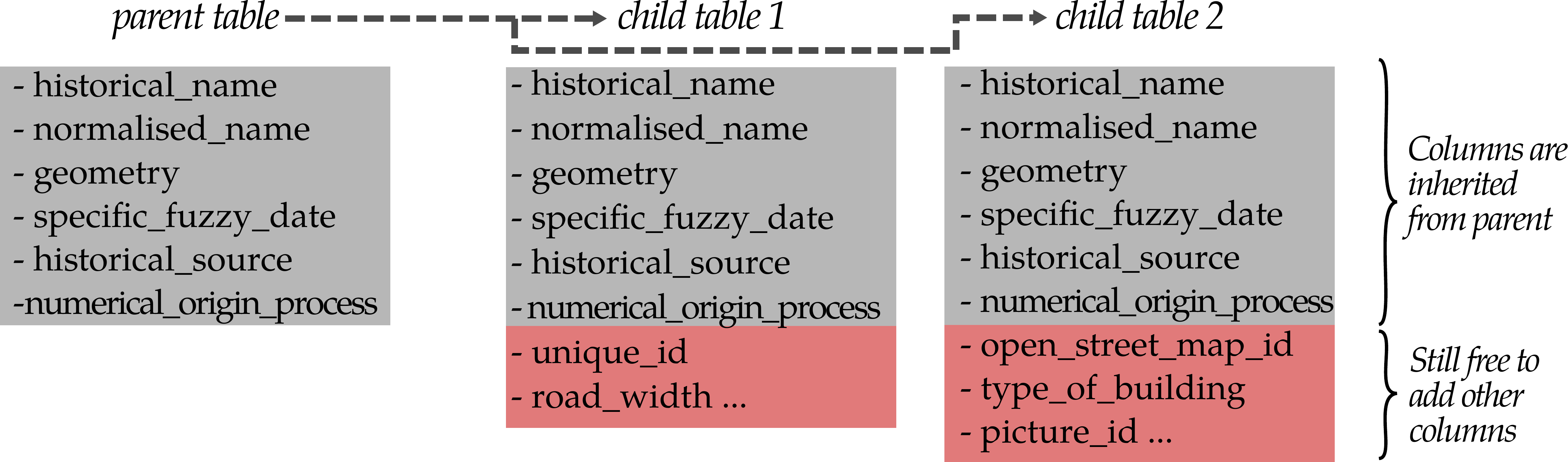}
			\caption{The table inheritance mechanism: a child table inheriting from a parent table inherits all the parent columns, and can also have its own. However, the parent table also virtually contains all the content of the child tables.}  
			\label{geocoding_method_table_inheritance}
		\end{center}
	\end{figure}

	We define a conceptual schema for geohistorical objects, which is based on two names, a source, a capture process, fuzzy dates and a geometry.
	This delineates the core of a generic geohistorical object.
	Yet this geohistorical object model is easily extendible using the table inheritance mechanism, an object-oriented design mechanism available in PostgreSQL (see Figure~\ref{geocoding_method_table_inheritance}).
	
	\paragraph{Table inheritance}
	The concept of table inheritance is simple, although slightly dissimilar to classic object-oriented inheritance.
	When a table $child$ is created as inheriting from a table $parent$, $child$ will at least feature the columns of $parent$, but can also contain other columns (provided there be no name/type collision).
	In our case, this means that a table of geohistorical objects will inherit from the main geohistorical object table, \emph{i.e.} will have all the core columns of geohistorical objects (names, sources, temporal aspect, spatial aspect), but can also have its own tailored column, providing the necessary flexibility. 
	Another key aspect of table inheritance is that when the $parent$ table is queried, the query will not only be executed the rows of the $parent$ table, but also on the rows of all $child$ tables.
	This means that all tables using the geohistorical object model will be virtually grouped and accessible from one table. This behaviour has no real equivalent in object oriented programming.
	
	\paragraph{\emph{Simulated inheritance of index and constraints}}
	The PostgreSQL table inheritance mechanism is however limited in some aspects, because constraints and index cannot be inherited.
	Constraints are essential, because they are used to guarantee that any geohistorical object will correctly use existing sources from the source tables ("historical\_source" and "numerical\_origin\_process").
	Indexes are also essential, because when using hundreds of thousand of geohistorical objects, they are needed to help speed up the queries.
	
	We index not only names, but all geohistorical object core columns (names, sources, temporal aspects, spatial aspects).
	We propose a registering function that the user can execute only once when creating a new geohistorical object table. This function then creates all the necessary indexes and constraints, and the appropriate inheritance. 
	
	\paragraph{\emph{Modelling a geohistorical object from the user's perspective}}
	The practical steps to create geohistorical objects are simple:
	\begin{enumerate}
		\item Add the historical source and numerical origin process in the source and process tables.
		\item Create a new table inheriting geohistorical objects and containing your additional custom columns
		\item Use the registering function with this table name
		\item Insert your data in the table.
	\end{enumerate}
	
	Please note that no disambiguation or comparison must be performed compared to other historical sources. Several historical sources with conflicting / duplicate information can co-exist without any problem.
	
	\subsection{The historical geocoder}
	
	In our method, geocoding something means finding the most similar geohistorical objects within the available gazetteers, which then provide the geospatial information.
	This approach relies on two key components: gazetteers of geohistorical objects, and a metric to find the best matches.
	This approach allows to perform geocoding in a broad sense, as it does not rely on a structured address (number, street, city, etc.), but rather on a non-constrained name.
	For instance the address ''Eiffel Tower, Paris'' is not structured, but would nonetheless be useful in our approach.
	
	\subsubsection{Creating geohistorical object gazetteers for geocoding}
	Geohistorical object gazetteers are key for geocoding. 
	These objects are extracted from historical maps and inserted into geohistorical object tables. Each table forms a gazetteer. 
	
\begin{figure}[htb!]
		\begin{center}
			\includegraphics[width=\linewidth,keepaspectratio]{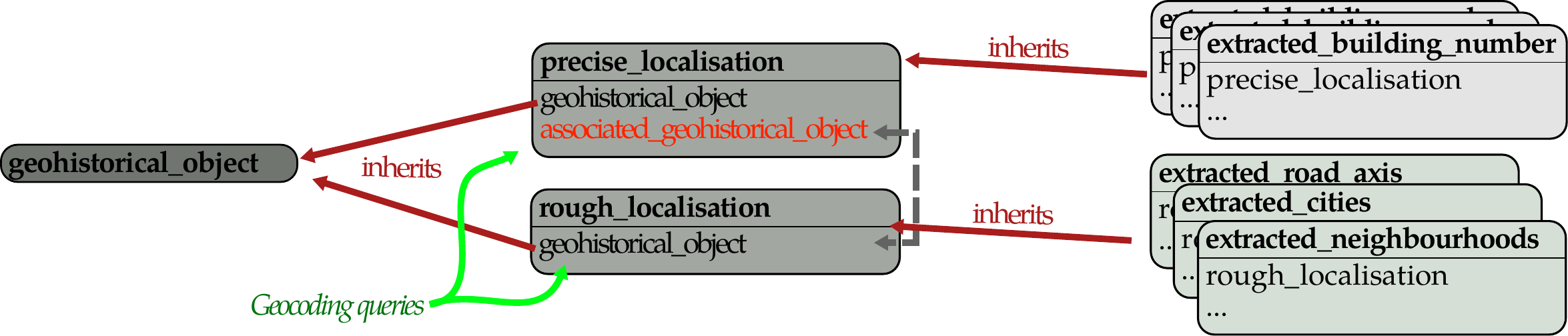}
			\caption{Geocoding table architecture. Two tables of geohistorical\_object are the support for geocoding queries. Because all extracted geohistorical object tables inherit from these two tables, they both virtually contain all the objects.}  
			\label{geocoding_method_geocoding_architecture}
		\end{center}
	\end{figure}

	\paragraph{\emph{Database architecture for geocoding}}
	Again, we use the PostgreSQL table inheritance mechanism.
	To this end, we create two tables dedicated to geocoding.
	Gazetteers tables which will be used in geocoding must inherit from these two tables.
	Table "precise\_localisation" is for geohistorical objects corresponding to postal addresses, \emph{e.g.} "12 rue du Temple, Paris".
	Table "rough\_localisation" is for road axis, neighbourhood and other coarse urban objects.
	We chose to have two separate tables for ease of use and performance.
	Geocoding queries are then performed on the two parent tables, but thanks to inheritance, these parent tables virtually contain all the gazetteers tables containing the actual geohistorical objects, as illustrated in Figure~\ref{geocoding_method_geocoding_architecture}.
	
	\subsubsection{Finding the best matches}
	\label{method.geocoding.matches}
	Once geohistorical object gazetteers describing precise and rough localisation are available, we geocode to find the best match between the input query and the objects.
	\paragraph{\emph{Concept}}
	
	We call the potential matches "candidates", and the problem is then to rank the candidates from best to worst.
	The user can choose how many candidates he wants, depending on the application.
	For an automated batch geocoding, the best match (top candidate) is optimal.
	For a human analysis of data, several matches may be more interesting (top 10 candidates for instance). 
	What qualifies as "best" depends on the user's expectations.
	We provide a number of metrics which can be combined by a user into a tailored ranking function.
	The function is expressed in SQL, with access to all postgres math functions. 
	We describe the available metrics and give examples of such functions.
	
	We note that recently other matching methods using probablity or machine learning have emerged (see \cite{Massey2017} for an evaluation).
	
	\paragraph{\emph{Example}}
	For instance when a user geocodes the address "12 rue de la Vannerie, Paris" in 1854, he or she may be more interested in geohistorical objects that are textually close (\emph{e.g.} a geohistorical object "12 r. de la Vannerie Paris", 1810), or maybe geohistorial objects that are temporally close (\emph{e.g.} "12 r. de la \emph{T}annerie Paris",1860).
	
	\paragraph{\emph{Metric: string distance $w_d$}}
	We use the string distance provided by the PostgreSQL Trigramm extension (pg\_trgm\footnote{\url{https://www.postgresql.org/docs/current/static/pgtrgm.html}}), which compares two strings of characters by comparing how many successive sets of 3 characters are shared.
	For instance "12 rue du Temple" will be farther away from "12 rue de la Paix" than from "10 r. du Temple".
	
	\paragraph{\emph{Metric: temporal distance $t_d$}}
	Both the address query and the geohistorical object are described by fuzzy dates.
	In order to compare such temporal information, we propose a simple fuzzy date distance that casts fuzzy dates into polygons. 
	The x axis is the time, and the y axis is the probability of existence of the object.
	Then the distance between two dates $A$ and $B$ is computed as shortest\_line\_length(A,B) + Area(A) - Area(A $\cap$ B).
	Note that this distance is asymmetric. 
	
	\paragraph{\emph{Metric: building number distance $b_d$}}
	To get building number distance, a function first extracts the building number both from the input address query ($b_i$) and from the geohistorical object ($b_d$).
	If $b_i$ and $b_d$ have same parity, the distance is $\mid b_d - b_i\mid$.
	If parity is different, the distance is $\mid\mid b_d - b_i\mid + 10\mid$.
	In France, building numbers generally have the same parity on each side of the street (\emph{e.g.} Left : 1,3,5,.. ; Right: 2,4,6..).
	We analysed current building numbers in Paris and determined that on average, given a building number $b_i$, the closest building number with a different parity has a 10 number difference.
	
	\paragraph{\emph{Metric: positional accuracy $s_p$}}
	Another way to rank the geohistorical objects is to use their positional accuracy.
	The positional accuracy of a geohistorical object is either the positional accuracy computed for this object when it is available, or the default positional accuracy of its geohistorical source.
	
	\paragraph{\emph{Metric: level of detail distance $s_d$}}
	Providing localisation information at different levels of detail, depending on user requirements is an important quality issue for our geocoder. 
	For instance, if the level of detail of the user's query data is the city, there is no need to perform a more precise geocoding.
	The user can therefore specify a target scale range $(S_l, S_h)$. Then, given a geohistorical object whose geometry is buffered ($geom_b$) with its spatial precision, the scale distance is defined by $least(\mid\sqrt{area(geom_b)}-S_l\mid, \mid\sqrt{area(geom_b)}-S_h\mid)$. 
	The formula $\sqrt{area(geom_b)}$ gives an idea of the geohistorical object's spatial scale.
	
	\paragraph{\emph{Metric: geospatial distance $g_d$}}
	The user may provide an approximate position for the area he is interested in.
	For instance, in France, cities "Vitry-le-Francois" (East) and "Vitry-sur-Seine" (near Paris) both exist, but are spatially very far apart.
	A user expecting results in the Paris area may provide a geometry (a point for instance) near Paris.
	The classic geodesic distance is then computed between the provided geometry and the candidate geohistorical objects.
	
	\paragraph{\emph{Example of matching function}}
	The different metrics can be weighted and combined depending on user needs.
	Equation~\ref{method.geocoding.matching.eq} provides an example favouring good string similarity, but not at the expense of a large temporal distance. 
	\begin{equation}
	100*w_d+0.1*t_d+10*n_d+0.1*s_p + 0.01*s_d+0.001*g_d
	\label{method.geocoding.matching.eq}
	\end{equation}
	
	\subsection{Collaborative editing of geohistorical objects} 
	The geocoding approach we have presented in the previous section works inside a PostgreSQL database.
	Given an input address and fuzzy date, plus a set of parameters, it returns the geohistorical objects that best matches the input.
	Yet the geocoding results are only as good as the gazetteers used (at best).
	The geohistorical objects within the gazetteers may be spatially imprecise, mistakenly named or simply missing.
	Given that the volume of geohistorical objects is large (for Paris, approximately 50 k building numbers per historical map), we created a collaborative platform which facilitates geocoding, result visualisation and geospatial object editing when necessary.
	To this end, we created a dedicated web application in order for collaborative edits to be made without having to install specific tools.
	The user can then edit both the position of the result, and the fuzzy date of the result. In fact, the user does not edit the sources, but actually edit a duplicate. This duplicates are stored and used by the geocoder as another gazetteer. 
	We do not try to merge/resolve several edits of the same address, as it is common for historical gazetteers, because there is no unique definition of an adress proper position. 
	By design, the quantity of data is then ever increasing, yet the great number of addresses (several hundreds of thousands), and the user profile (expert or historians) limit this potential problem.
	
	\begin{figure} [htb!]
		\begin{center}
			\includegraphics[width=\linewidth,keepaspectratio]{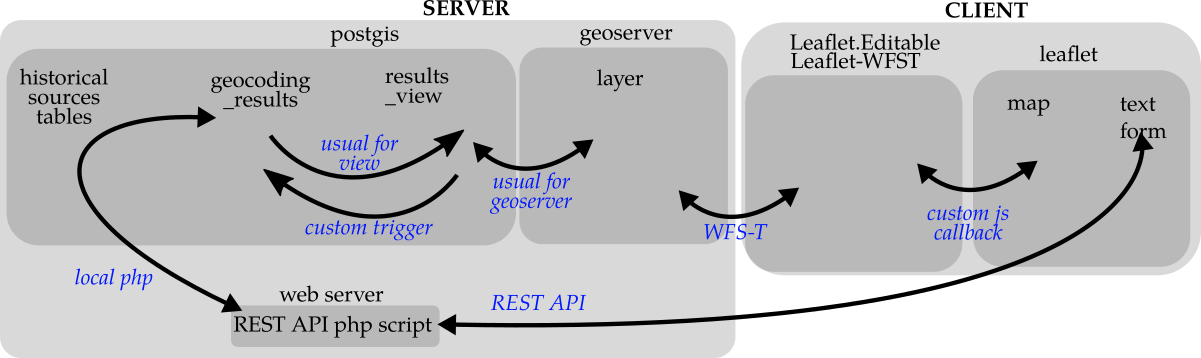}
			\caption{Conceptual architecture for interactive display and edit of geocoding results. The stack contains only standard components. The use of WFS-T and REST standard protocoles makes the change or customisation of some components easier.}  
			\label{geocoding_method_edit_architecture}
		\end{center}
	\end{figure} 
	
	\subsubsection{About collaborative editing}
	Given the complexity of calibrating automatic extraction tools on specific maps and their relative reliability, the collaborative digitisation of vector objects from maps is a safe alternative.
	For instance, we used such an approach in order to extract the main feature of the Cassini maps ($18^{th}$ century France)~\cite{Perret2015}.
	Furthermore, the results of the collaborative extraction of features can then be used to test, calibrate or train automatic extraction algorithms.
	The collaborative editing paradigm used is, however, somewhat different from the classic one, a-la Open Street Map, which has also been used in several historical mapping projects.
	In the classic paradigm, users are asked to input and correct well defined historical data, such as building footprints or toponyms, based on a map where this information appears unambiguously. 
	The end goal is then to create authoritative, complete, and coherent gazetteers.
	This requires a large work over the users' inputs, using strategies such as vote and anonymous check to ensure the gazetteer quality.
	
	Our approach, however, introduces a much simpler collaborative editing process, whereby the users are not tasked with creating an authoritative gazetteer, but rather create their own version tailored after their specific needs. In our model, users never edit original data, but instead create their own geohistorical objects.
	
	For instance, a user might geocode the address "12 rue du Temple, Paris; 1856". The geocoded result (a point for instance) might be drawn from an available gazetteer created from a historical map.
	Such a point position may be not accurate enough for the user, and he or she may decide to correct it. 
	While correcting the point, the user is not editing the gazetteer but just adding a new geohistorical object into a new gazetteer which represents this user's edits.
	Such edits can subsequently be modified by the same user, and may appear in other users' geocoding results.
	However, no step is performed to aggregate user edits.
	The reason is that unlike a building footprint, an address position is not something that is well defined.
	Different historians may use different definitions of what an address position should be.
	One might require for addresses to be centred in the building, others for them to be at the front of the building, etc.
	
	The necessity not to aggregate user edits is even more obvious when considering the address date.
	Several historians may use different sources to date an address, leading to the creation of several geohistorical objects representing this address at different time periods.
	
	A potential issue would be data build-up, as each edit may introduce new data.
	Such issue impact is greatly reduced for several reasons. The first is that the goal is not to create a reference gazetteer. As such, merging is not required per se. The second is relevant to scale. Just for the city of Paris, there are hundreds of thousands of building numbers, with very frequent changes. Thus, the chances that a single address gets edited many times by many different users is low. The third reason is more theoretic. 
	The increasing amount of data is actually a useful feature.
	Several edits by several users will result in several edited geohistorical objects being added to a specific gazetteer (the user-edit gazetteer). In turn these results will be used by the geocoder, which will enrich a future user experience. Indeed, a future user geocoding the same address would be able to see all the edited version, and thus chose the more appropriate one according to his needs.

	\subsubsection{Collaborative editing architecture}
	Figure~\ref{geocoding_method_edit_architecture} outlines the architecture used for collaborative editing.
	\paragraph{\emph{Architecture}}
	\label{method_geocoding_edit_architecture}
	The heart of the architecture is a PostgreSQL database server, which contains the geohistorical object gazetteers to be used for geocoding as well as the geocoding function.
	A web server can geocode addresses and return results via a REST API.
	However, the web server has another option wherein the results are not returned, but instead written in a result table along with a random unique identifier (RUID).
	The RUID is then the key that allows the display and editing of the results.
	To this end, a geoserver can access (read and edit) the result table via the WFS-T protocol.
	A web application based on Leaflet then acts as a user interface to display and edit the results via the geoserver.
	
	\paragraph{\emph{Persistence of geocoding results and edits}}
	\begin{figure} [htb!]
		\begin{center}
			\includegraphics[width=\linewidth,keepaspectratio]{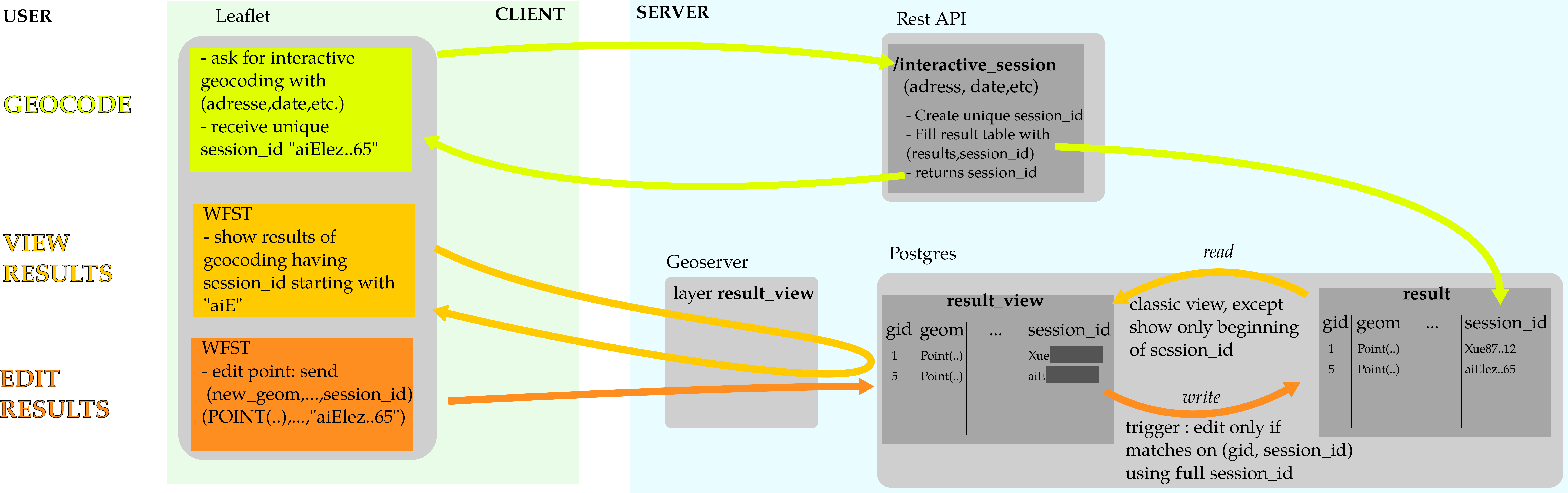}
			\caption{Collaborative display and edit is achieved through a mix of standards (REST , WFS-T) and custom solution (triggers) that enable the sharing of a basic public/private key.}  
			\label{geocoding_method_edit_ruid}
		\end{center}
	\end{figure} 
	The architecture that allows persistence of results is illustrated in Figure~\ref{geocoding_method_edit_ruid}.
	When using the RUID mechanism, each geocoding result (that is the found geohistorical object from the gazetteers) is associated to this RUID.
	The user therefore has permanent access to its results, regardless of the computer session or browser cache issues.
	
	For edits, a specific mechanism is used.
	The user does not directly edit the result table, as he could potentially edit other peoples' results.
	Instead, the user edits a dedicated result\_view which acts like a bouncer.
	It allows one to edit only if the edit is occurring on a row that has the user's RUID.
	Of course, user edits of geospatial objects do not affect the source data, for tracking purposes.
	
	Instead, a new user edit automatically creates an edited copy of the geohistorical object in a dedicated table "user\_edit\_added\_to\_geocoding" which is a gazetteer and is used by the geocoding process.
	The edited geohistorical objects are inserted in this table.
	The objects retain their "historical\_source", but their "numerical\_origin\_process" is changed to properly document the fact that they are the result of collaborative editing.

	\subsubsection{Collaborative editing user interface} 
	We consider that building an efficient user interface is very important for historical geocoding. 
	In particular, many end users are specialised in history rather than computer science, and thus an easy access to geocoding is essential.
	All our interfaces are web-based for maximum compatibility.
	We propose three interfaces where results are shared.
	
	\begin{figure} [htb!]
		\begin{center}
			\includegraphics[width=\linewidth,keepaspectratio]{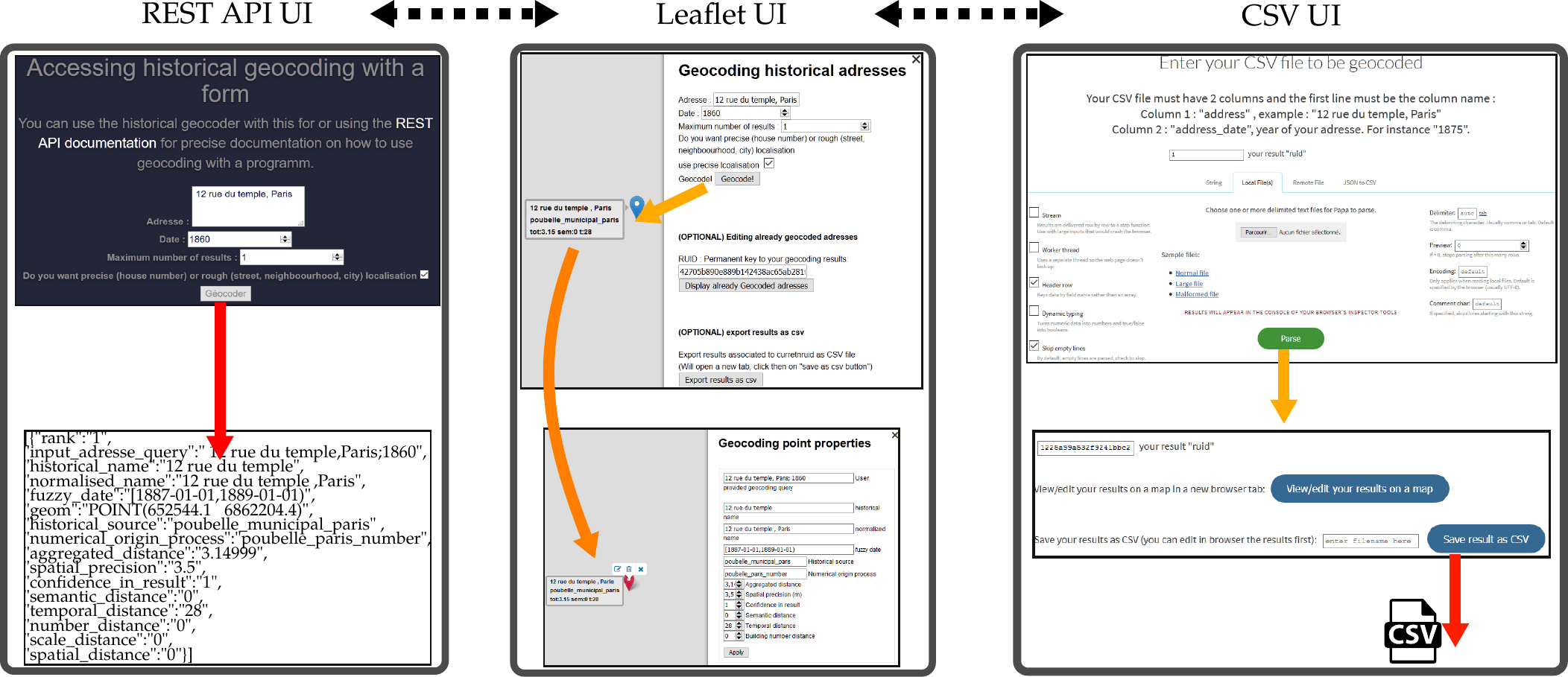}
			\caption{Various Web User Interfaces that can be used to access the proposed historical geocoding tool.}  
			\label{geocoding_method_UI}
		\end{center}
	\end{figure} 
	
	\paragraph{Interface for a REST API.}
	The simplest interface we propose is a form that helps build the necessary REST API parameters.
	Indeed, a REST API works via URL containing precise parameters, and it can be tedious to manipulate.
	For instance: \\
	\url{http://api.geohistoricaldata.org/geocoding?address=12 rue du temple&date=1850&precision=true&maxresults=1}\\
	This interface is designed to be used in an automated way, for batch geocoding.
	
	\paragraph{Interface for batch geocoding via CSV files.}
	In our experience, historians often work with spreadsheet files, where each line will be a potential historical object, along with an address and a date.
	To facilitate the geocoding of these addresses, we propose a User Interface which reads Coma Separated Value (CSV) files (a standard spreadsheet format) and geocode the address and date they contain.
	This interface is built around the PapaParse\footnote{\url{http://papaparse.com}} Javascript framework. 
	The geocoding results can then either be downloaded as a CSV file, or displayed and edited in a web application.
	
	\paragraph{Interface for display and edit of results.}
	The most complex interface we propose is based on the Leaflet\footnote{\url{http://leafletjs.com}} Javascript framework. 
	There, the user can geocode an address, or use an address which has already been geocoded via the RUID mechanism (see Section \ref{method_geocoding_edit_architecture}), be it from previous sessions or from geocoded CSV files.
	The geocoding results are displayed on top of a relevant historical map, and can be edited.
	Users can edit result geometry as well as result names (historical and normalised).
	We stress that although such edits are stored into the database, and used by further geocoding queries, they do not, by design affect source data.
	
	
	\section{Results}
	Several experiments are performed to validate our approach.
	First, the geohistorical model is used to integrate objects extracted from historical maps from the 19\textsuperscript{th} century for the city of Paris, and the current OpenStreetMap road axis and building numbers for Paris city surroundings.
	Road axis, building numbers, and neighbourhoods are successfully integrated to the geocoder sources.
	Multiscale geocoding of dozens of thousand of historical addresses is then performed. Addresses are extracted manually by historians and extracted automatically through an automatic process.
	For one of the datasets, a historian manually corrects the automated geocoding results, so as to evaluate the quality of our method.
	Last, the collaborative editing of geohistorical objects is evaluated in two scenarios: analysis (several results for one address), and edit (efficiency of check/edit top results for several addresses). 
	
	\subsection{Geohistorical objects sources}
	Three main historical sources of geohistorical objects are used to build gazetteers and perform geocoding.
	The first two are historical maps of Paris from the 19\textsuperscript{th} century.
	These maps are georeferenced, then street axis (and possibly building numbers) are manually extracted.
	The third historical sources are road axis and building numbers for Paris surroundings extracted from current Open Street Map data.
	\begin{figure} [htb!]
		\begin{center}
			\includegraphics[width=\linewidth,keepaspectratio]{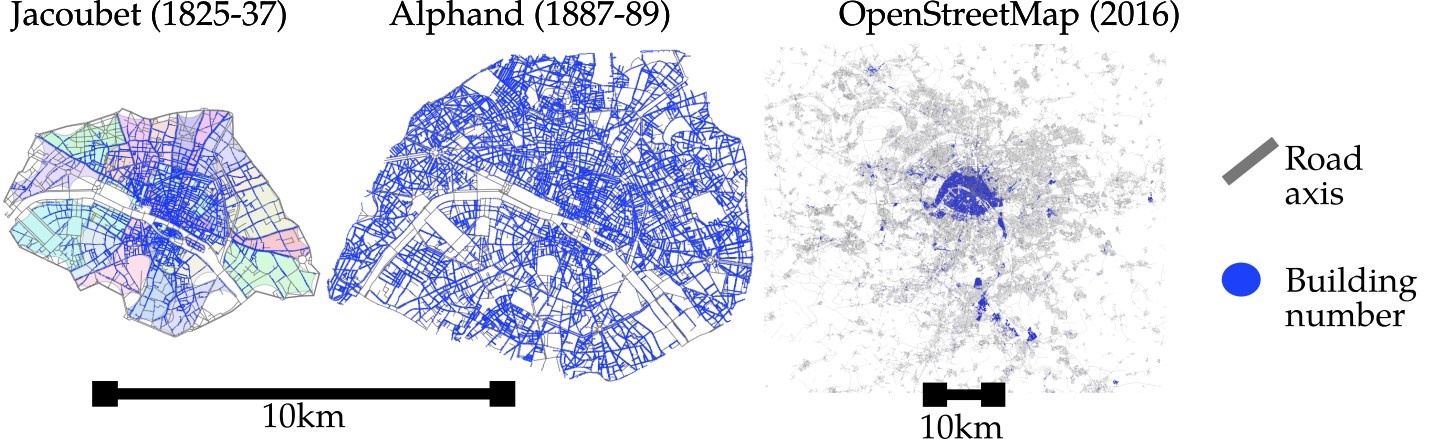}
			\caption{An overview of some of the geohistorical objects used for geocoding. These have mostly been extracted manually, often in a collobarative fashion, using various tools.}  
			\label{result_geohistorical_sources}
		\end{center}
	\end{figure}
	
	\subsubsection{Historical maps used}
	Two major French atlases of Paris from the 19\textsuperscript{th} century are integrated as geohistorical sources. 
	The first one is the "Atlas municipal de la Ville, des faubourgs et des monuments de Paris"\footnote{Municipal atlas of the city, suburbs and monuments of Paris.} created at the scale of $1:2000$ between 1827 and 1836 by Theodore Simon Jacoubet, an architect who worked for the municipal administration of Paris.
	The second atlas is the 1888 edition of the "Atlas municipal des vingts arrondissements de la ville de Paris"\footnote{Municipal atlas of the 20 districts of Paris}.
	For legibility reasons, we refer to the first atlas as the "Jacoubet atlas" and to the second as the "Alphand atlas"\footnote{Named after Jean-Charles Alphand who wasthe director of the department of public works of Paris at that time.}. 
	The Jacoubet atlas depicts a city standing between the housing development following the sale of properties which had been confiscated during the French Revolution and the majors changes in the urban structure arising from the emergence of the fist train stations in 1837-1840 and the so-called Haussmannian transformations. 
	
\begin{figure} [htb!]
		\begin{center}
			\includegraphics[width=0.8\linewidth,keepaspectratio]{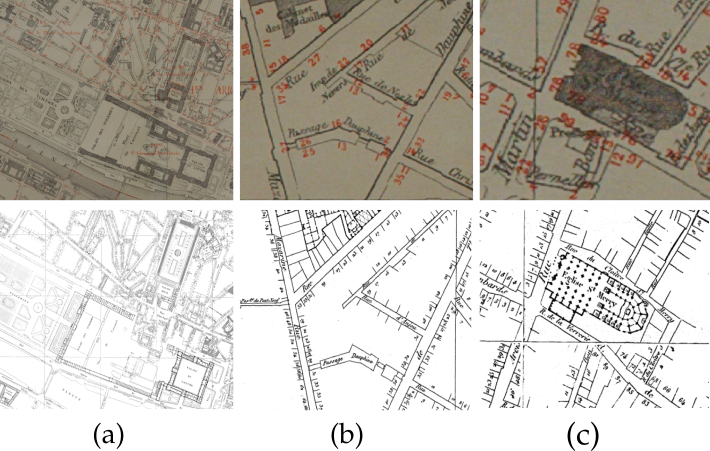}
			
			\caption{Samples of the georeferenced Alphand Atlas (2\textsuperscript{nd} row) and Jacoubet Atlas (1\textsuperscript{rst} row) at different scales: district (a) and urban islet (b). Column (c) shows how buildings are portrayed in the maps.}
			\label{figure:extract_jacoubet_alphand_atlases}
		\end{center}
	\end{figure}

	The Alphand atlas is a portrayal of Paris on a scale of $1:5000$, erected after most of the Haussmannian transformations (major rework of Paris urbanism in the 19\textsuperscript{th} century) had been made and after the city was merged with 11 of its neighbouring municipalities in 1860.
	Both atlases contain large scale views of Paris, separated in several sheets (54 and 16 respectively) and portray the urban street network with the name of each street, as well as public and religious buildings (see Figure~\ref{figure:extract_jacoubet_alphand_atlases}).
	In addition, the house numbers are specified for most of the streets in the city, although the Alphand atlas pictures only the numbers at the start and end of each street section.
	Both atlases are also built upon a triangulation canvas which covers the entire city, enabling us to expect a high positional accuracy of the geographical features they contain.
	
	The two atlases are georeferenced using the grids drawn on the maps, which are aligned on the Paris meridian, as a pseudo-geodetic object to identify feature pairs.
	The dimensions of the grid cells also appear on the maps, allowing us to reconstruct the grids in a geographic reference system.
	The Lambert I conformal conic projection was chosen to georeference the maps. It uses the Paris meridian as the prime meridian and relies on the NTF (Nouvelle Triangulation Française) geodetic datum.
	The main advantage of this projection is that it is locally close to the planar triangulation of Paris used in the atlases.
	The projection of the maps can thus reasonably be approximated by the Lambert I projection, making the reconstruction of the grids in the target coordinate reference system straightforward.
	In addition, since both maps have a high scale and are reliable because they are official maps with high positional accuracy, we used rubbersheeting as the geometric transform model.
	The georeferencing process applied for each atlas was the following:
	\begin{itemize}
		\item reconstruct the meridian-aligned grid with Lambert I coordinates;
		\item in each sheet, mask the non-cartographic parts out (cartouche, borders, etc.); 
		\item for each sheet, set pairs of ground control points at each intersection between the vertical and horizontal lines of the grids in the map and in the reconstructed grid;
		\item transform each sheet with a rubbersheeting transform based on the ground control points previously identified on the grids.
	\end{itemize}
	
	Based on these atlases, vector road axis are manually drawn and the road name inputted for the Alphand map. The building numbers at the beginning and end of each street segment are also inputted.
	For Jacoubet, the building numbers from a previous map (Project Alpage, Vasserot map, \cite{Noizet2013}) are adapted to fit the Alphand map.
	Multiple series of successive checking and editing are performed using ad hoc visualisation and tools.
	
	For Alphand, building numbers are then generated based on available information (for each street segment, for each side, beginning and ending number) by linear interpolation, and an offset.
	The size of the offset is estimated by using current Paris road width when the road has not changed too much.
	Overall, the process is quite similar from \cite{Dhanani2016} work.
	
	\subsubsection{Other geohistorical sources}
	The presented system accepts any data that conforms to the previously introduced geohistorical object model. As such, we also introduce data from OpenStreetMap, dated from 2016.
	As a comparison, \cite{Carrion2016} use uniquely current data to geocode medival places.
	We use the version of the data which has been transformed to be used by the Nominatim geocoder.
	Custom scripts extract road axis and building numbers, which are then converted into a geohistorical object table.
	Spatial precision is estimated after a short analysis of the positioning of a few Paris addresses.
	Using Open Street Map addresses highlights several of the possibilities the proposed method has to offer. 
	First, our geocoder can work seamlessly with historical and modern data. The user can choose which type of data to use by placing more or less importance on temporal distance. 
	Furthermore, the OSM addresses may act as a safety net for addresses that do not appear in any other historical gazetteer. Last, using modern data may be of interest for further address evolution analysis.
	We stress that for the geocoding system, the OSM data is just another geohistorical data set that happens to be dated from around 2016. Other modern address datasets could similarly be added.

	\subsection{Geocoding of Historical datasets}
\begin{figure} [htb!]
		\begin{center}
			\includegraphics[width=\linewidth,keepaspectratio]{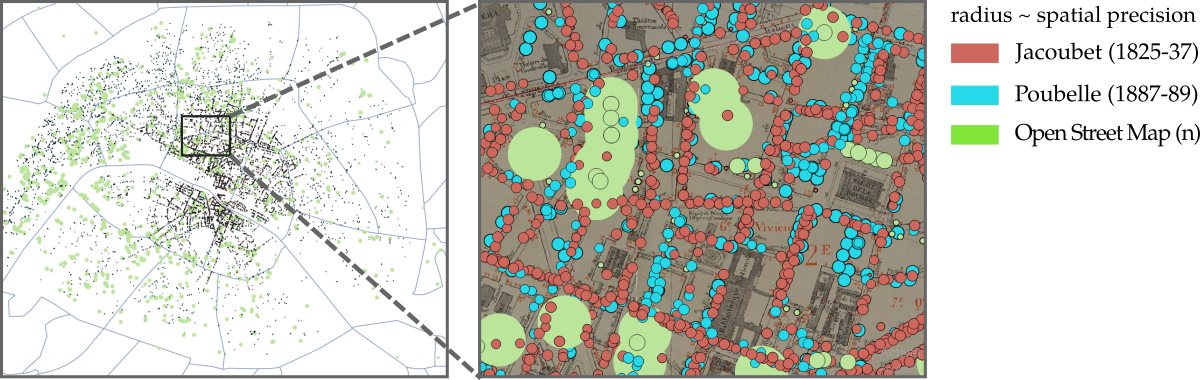}
			\caption{Illustrations of all the historical textual address data sets that have been geocoded. Result size is proportional to spatial precision.}  
			\label{results_geocode_alldatasets}
		\end{center}
	\end{figure}

	One of the end goal of our geocoding tool is to be useful for historians.
	Therefore, we contacted several historians working on 19\textsuperscript{th} century Paris.
	They had been collecting historical textual addresses associated to a person or business for their own research, by manually browsing hundreds of archive documents.
	Overall, the collected textual addresses are of good quality (being hand collected), although they sometimes contain errors and abbreviations.
	We imported their data into the geocoding server and geocoded the provided addresses (i.e. associated matching geohistorical objects from the gazetteers).
	Figure~\ref{results_geocode_alldatasets} shows an extract of the thousands of geocoded addresses, while table~\ref{results_geocode_alldatasets_table} gives an overview of the number of successes and timing.
	
	\begin{table}
		\begin{tabular}{|c|c|c|c|}
			\hline 
			Dataset name& input addresses  &  response rate (rough) & secs/1000 addresses\\ 
			\hline 
			South Americans & 13991 & 13743 (250) & 138  \\ 
			\hline 
			Textile & 5777 & 5688 (16) & 135 \\ 
			\hline
			Textile 2 & 3070 & 3053 (2) & 110 \\ 
			\hline
			Artists accommodations & 13907 & 10215 (2955) & 244\\ 
			\hline
			Health administrators & 1887 & 1698 (171) & 316 \\ 
			\hline
			Belle epoque (0.3) & 6467 & 3880(337) & 280 \\ 
			Belle epoque (0.5) & 6467 & 6000 & 351 \\ 
			\hline
		\end{tabular} 
		\caption{For all textual address historical data sets, how much addresses have been geocoded and how long it took.}  
		\label{results_geocode_alldatasets_table}
	\end{table}
	\subsubsection{Manually collected dataset}
	
	\noindent \textbf{South Americans:} Collection of South America immigrants living in Paris in 1926, manually input from census, collected by Elena Monges (EHESS).\\
	\textbf{Textile:} Collection of professionals of textile industry in Paris, manually input from the "Almanachs du Commerce de Paris", from 1793 to 1845, collected by Carole Aubé (EHESS).\\
	\textbf{Artists accommodations:} Textual addresses of artist studios and artists accommodations between 1791 and 1831, collected by Isabelle Hostein (EHESS) to study their impact on Paris' development.\\
	\textbf{Health administrators:} Addresses of public health and hygiene administrators in Paris between 1807 and 1919 (\cite{Gribaudi1999}), collected by Maurizio Gribaudi and Jacques Magaud (INED-EHESS).
	
	\subsubsection{Belle Epoque}
	The Belle Epoque dataset is different from the previous one because it has been automatically extracted from directories of Paris financial societies between 1871 and 1910.
	Directories are books referencing company addresses (as well as names and other information).
	The process of automatic extraction is in itself complex (Project Belle Epoque, \cite{Lazzara2011}), and is out of scope of this article.
	We can only provide a brief description below.  
	
	First, each page of the directories of Paris for specific years has been photographed. 
	Pictures are then straightened, and information is extracted via an OCR software which has been configured for the directory's specific layout.
	Further rule-based processing parses the text into address fields. 
	As a result of this automatic process, the quality of addresses is often significantly lower than manually edited addresses (characters may be wrong, other textual fields may have bled into the address field, etc.).
	Therefore, we test two settings by allowing a greater maximum string distance from 0.3 to 0.5 (over 1).
	
	\subsection{Manual editing of the geocoding results for evaluation}

	\begin{figure} [htb!]
		\begin{center}
			\includegraphics[width=\linewidth,keepaspectratio]{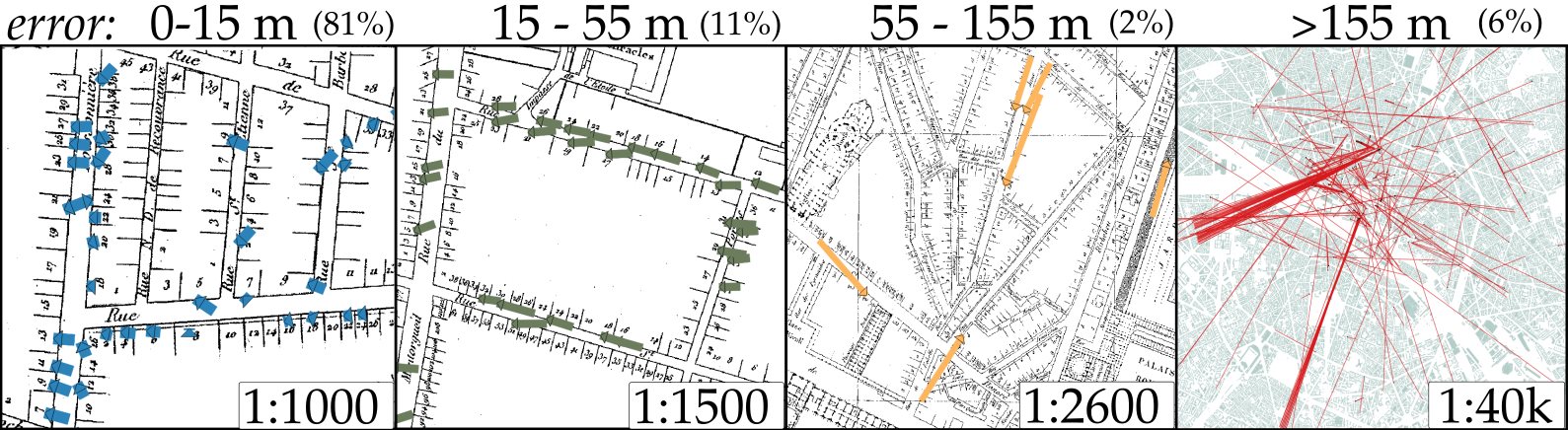}
			\caption{A textual address data set was geocoded, then manually corrected by an historian. A historian manually corrects the geocoded addresses. We plot the segment between the geocoded address and the corrected address, and analyse the results based on the magnitude of the spatial error. }  
			\label{results_geocode_analysis}
		\end{center}
	\end{figure} 
	
	For one of the data sets (Textile 1 and 2), the historian manually corrected the positions of the geocoding results (i.e. the positions the goecoding associated to input textual addresses).
	The corrected positions now form a ground truth dataset (associating textual addresses and corrected positions).
	We geocode this ground-truth dataset again and analyse the results, namely to try and understand the accuracy of the geocoder.
	
	The segment between address point resulting from automated geocoding and address point after manual editing (ground truth) is plotted. 
	Results are presented in the table \ref{results_geocode_evaluating} and in Figure~\ref{results_geocode_analysis}
	We classify the results based on the length of this segment (\emph{i.e.} the error in metre generated by the geocoding method).
	\begin{itemize}
		\item   When the edit moves the address point by less than 15 metres, we can consider that the edit is mostly about small moves, for instance centring the point on the building limit.
		\item Between 15 and 55 meters, the correct street has been found, but the building numbers are slightly misplaced (a few numbers).
		\item  Between 55 and 155 meters, the street is correct in most cases, but the building numbers are far from their correct position.
		\item  Above 155 meters, streets are mostly wrong. 
	\end{itemize}
	
	We stress that given Paris buildings' average size and the lack of precise definition of an address (is it the position of the door, the center of the building, etc.?), results up to 55 metres (>92\% of dataset) could be considered as very close to ground truth. 
	
	\begin{table}
		\begin{tabular}{|c|c|c|c|c|c|}
			\hline 
			dist. (m) & \% & avg(agg) & avg(sem) & avg(tempo) & main edit cause (subjective)\\ 
			\hline 
			0 - 15   & 81 \% & 9.4 & 0.07 & 19.5 & moving point on building limit\\ 
			\hline 
			15- 55   & 11 \% & 12.4 & 0.09 & 27.2& small numbering editing (same street)\\ 
			\hline
			55  - 155  & 2  \% & 23.7 & 0.14 & 41.2 & large numbering editing (same street)\\
			\hline 
			155 - 7.2k & 6  \% & 26.9 & 0.18 & 49.1 & editing street \\
			\hline
		\end{tabular} 
		\caption{Evaluating the error of geocoded results, via the dist. (geographic distance) of edit (in metres), the percentage of the total 8823 addresses, the average aggregated distance score, the average string distance $w_d$, the average temporal distance $t_d$, and the subjective most common edit reason we encountered while browsing the data}  
		\label{results_geocode_evaluating}
	\end{table}
	
	\subsection{Collaborative editing}
	We propose several User Interfaces for easy geocoding, and collaborative editing of the geocoding results.
	We informally tested the interfaces and found that they facilitated geocoding, especially for the batch mode.
	We also test the collaborative editing in two use cases. 
	In the first use case, a specialised user geocodes a single address and displays the top 3 corresponding results. The user is an expert and his or her goal is both to geocode an address and assess the reliability of the result at the same time.
	In the second use case, a user batch geocodes several addresses (30), looking at the best result for each address. The user then displays the results on the map and checks/edits the adresses.
	Please keep in mind that edits never change the gazetteers, but rather create new geohistorical objects.
	
	\begin{figure} [htb!]
		\begin{center}
			\includegraphics[width=\linewidth,keepaspectratio]{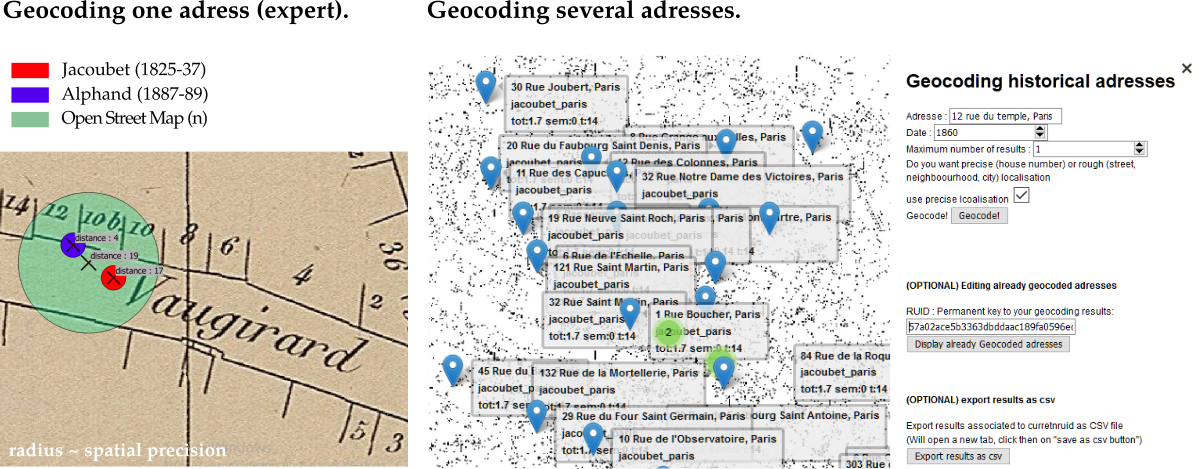}
			\caption{Two use cases: First use case, an expert geocodes an address and analyses the top 3 results to assess the reliability of the result. Second use case: a user batch geocodes 30 addresses ( 1 result per address) in Paris and checks/edits the results.}  
			\label{result_interactive_scenarios}
		\end{center}
	\end{figure}
	
	\subsubsection{Use case 1: top 3 results for one address}
	Using the web application, we geocode the address "10 rue de Vaugirard, Paris" for the date 1840, and ask for the top 3 results, as shown in part one of illustration \ref{result_interactive_scenarios}.
	A matching building number geohistorical object exists in the three gazetteers extracted from the three maps.
	Based on the results, we can safely assume that this building number has not changed during the last 2 centuries.
	
	\subsubsection{Use case 2: batch geocoding of 30 addresses and check/edit}
	In this use case, a regular user is to check/correct 30 random addresses from the Jacoubet map using the web application.
	The task is performed quickly, checking and editing each address is a matter of seconds. The main time consuming task is the loading of the background historical map, due to unfortunate hardware limitations. The edit speed seems to be on par with a desktop based editing solution (using QGIS).
	
	\section{Discussion}
	
	\subsection{Genericity}
	\label{subsection:genericity}
	Reaching a more generic geocoding service is important if we want to make it usable in other contexts and to profit from the various sources of knowledge on past spaces.
	\subsubsection{Geohistorical sources and data}
	\paragraph{\emph{Using external resources from the Web of data as new sources}}
	Beside features representing address points and streets, georeferenced features of other types could be benefit from the geocoding service.
	As a matter of fact, people often refer to places of interest, such as famous buildings, monuments like statues or fountains or even identified neighbourhoods to describe their position in space.
	We are thus considering adding data about places of interest to improve our geocoding service.
	Like the data used to build the geocoder, this data could be gathered from ancient maps.
	It may also originate from existing gazetteers and knowledge bases published on the Web of data, such as DBpedia\footnote{\url{http://wiki.dbpedia.org/}}, Yago\footnote{\url{http://www.mpi-inf.mpg.de/departments/databases-and-information-systems/research/yago-naga/yago/\#c10444}}, the Getty Thesaurus of Geographical Names \footnote{\url{http://www.getty.edu/research/tools/vocabularies/tgn/}} or the gazetteer of place names published by the French National Library\footnote{\url{http://data.bnf.fr/}}.
	\paragraph{\emph{Widen the spectrum of cartographic sources}} 
	We draw from Jacoubet and Alphand maps, yet there are several other maps to be used, dating from the end of the 19\textsuperscript{th} century, and in the beginning of the 20\textsuperscript{th} century.
	From the beginning of the 20\textsuperscript{th} century, the Paris city administration produced one map per year. 
	Of course, the main improvement direction would be to add maps of other cities/countries!
	For France at least, major cities have often been mapped starting in 1900.
	
	Before the beginning of 19\textsuperscript{th} century, the address system was very different in Paris.
	In the mid 18\textsuperscript{th} century, the address system consisted in each building having a specific name (no number, no notion of street name) in its neighbourhood.
	Our geocoding process has also been designed to integrate this type of address system, but it has not been tested yet. 
	More generally, this type of indirect localisation very closely resembles the field of web of knowledge.
	
	\paragraph{\emph{Diversity in geohistorical object natures}}
	In this article, several types of geohistorical objects were used for geocoding: building numbers, streets axis, neighbourhoods.
	Other datasets were investigated as well, such as the city limits collaboratively extracted from the Cassini maps by the Geo Historical Data project~\cite{Perret2015}.
	In fact, a compiled version of city limits (GeoPeuple project~\cite{Plumejeaud-Perreau2014}) from 1793 to 2010, created by EHESS, has also been tested.
	But building cadastres could also be integrated so as to have a building layout associated to an address rather than a point, which would solve an old problem of address points.
	Indeed, there is currently no consensus as to where a building number address point should be positioned: on the entry door, on the letter box, etc.
	In some cases, there is more precise data available, providing the layout of apartments in buildings, which is very exciting. 
	
	\subsubsection{Genericity in usages}
	\paragraph{\emph{Named Entity Linking}}
	As previously mentioned in section~\ref{subsection:genericity}, people often refer to place names to describe their position in space.
	The task of retrieving place names in a gazetteer or in a knowledge base, also known as (Spatial) Named Entity Linking or toponym resolution, is a widely used way of disambiguating mentions of spatial named entities extracted from texts by means of natural language processing approaches for information retrieval, information extraction or document indexing purposes~\citep{shen2015entity}.
	As we plan to upgrade our geohistorical database with data about places of interest, we also have to adapt our geocoding service in order to make it retrieve reference data stored in the database and corresponding to place names mentions proposed by the users.
	Spatial Named Entity Linking implies solving issues related to places names inherent ambiguity~\citep{overell2011}, such as the fact that a place may have several names or the fact that several places may be designated by the same name.
	For each spatial named entity mention to be disambiguated, unsupervised state of the art approaches first select candidates from the gazetteer based on character string similarity.
	Then, they introduce additional criteria in order to decide which candidate is the best reference for a given place name, usually taken from the textual context of the mention~\citep{Mih2007,Hachey2013}.
	In cases where textual context is very limited, like in tweets or location descriptions extracted from directories, this step of candidate ranking is even more challenging \citep{zhang2014geocoding}. 
	
	\paragraph{\emph{Analysis tool of the cartographic sources content}}
	It is interesting to look at which historical sources were the most used for geocoding, although the historical sources are chosen based on a complex ranking function.
	If we take the example of the over 10k geocoded addresses from the "Artists accommodations" dataset, we could expect all of the results to be drawn from the Jacoubet map, as the dataset is between 1793 and 1836, and the Jacoubet map is also in this range.
	Yet, analysing the results shows that if Jacoubet was used for 80\% of the addresses, Alphand was used for 15\%, although the map was issued 30 years after Jacoubet.
	More surprisingly, the OpenStreetMap current data is still used for 5\% of addresses, although it is about 2 centuries older than the dataset.
	
	Similar analyses of other datasets show that all maps are always used, with of course a focus on the temporally closest map.
	Interestingly, these results are in agreement with similar work as presented in~\cite{Costes2016}, chapter 4, where a prototype of multi-temporal geocoding is proposed.
	The approach shows that for different datasets, all references maps (Jacoubet, Alphand and BDAdresse (2010)) are used, with proportions depending on the parameters at play and the weight of each criteria.
	We think that these results are explained by the fact that historical maps miss some information, contain errors, and do not have the same geographical coverage.
	
	\subsection{Quality of the geocoding}
	\subsubsection{Increasing the quality of the gazetteers}
	\paragraph{\emph{Collaborative enrichment}}
	We propose several easy ways to use the geocoding capacities through web based User Interfaces.
	As we put prototypes forward, the experiments are merely proofs of concepts for the moment.
	For a real validation, a complete user study would be required, which is outside the scope of this article.
	
	\paragraph{\emph{Cross-referencing historical maps}}
	One way to improve quality of available historical data is to use advanced cross-referencing.
	Indeed, the process of linking and merging similar data from heterogeneous datasets, which is called data conflation, enables to transfer information from one feature to the another, and may thus bring additional knowledge about data imperfections without using ground truth data which are non-existent for geohistorical data.
	For instance,~\cite{Costes2015, Costes2016} proposed an aggregated spatio-temporal graph to merge and confront historical road networks.
	This process can reduce data heterogeneity and allow the detection of aberrations such as toponymic or numbering errors, as well as doubtful temporal trajectories of objects such as short disappearances, thereby leading to better data quality.
	Advanced cross-referencing also allows for the construction of a genealogy of addresses by considering temporally linked addresses, which can deal with toponymic evolution or changes in addressing systems or numbering of buildings, thus paving the way for better spatio-temporal geocoding results.
	
	\subsubsection{Communicating the reliability of a geocoding}
	\paragraph{\emph{Geocoding qualification and quality measures}}
	Modern geocoders are evaluated by how often they find a localisation, and how precise their returned localisation (see~\cite{Zimmerman2007} for instance).
	The first criterion illustrates the geocoding algorithm's ability to retrieve an address as well as the gazetteer's exhaustiveness.
	The second criterion refers to the positional accuracy of the gazetteer.
	Using such quality evaluation measures which encompass both the algorithm results and the gazetteer completeness makes the evaluation of their respective quality impossible.
	In the field of named entity linking, however, distinct quality evaluation measures have been proposed for to the entity retrieval algorithm, such as the measures introduced by~\citep{Hachey2013} and completed by~\citep{csimqBrandoFG16}, and for the reference knowledge base (see~\citep{zaveri2016quality} for knowledge bases general quality measures and~\citep{brando2016} to evaluate the fitness of some knowledge bases for a given named entity linking task).

	\paragraph{\emph{Geovisualisation}}
	The prototype of graphic user interface we put forward could be improved in several ways.
	The goal would be to efficiently provide the user with information about the quality of geocoding, and the context of results.
	First, the size of the point displayed to represent the result could be proportional to estimated spatial precision.
	This would help to visually assess relevant information.
	Second, the result could be colour-coded to represent the temporal proximity with he input date. 
	In a similar way, when multiple results are proposed, a time slider would be most useful to graphically distinguish result candidates from one another.
	Third, the background historical map displayed in the prototype is currently set.
	Yet, the most appropriate background map could be automatically displayed based on the input address dates provided by the user.
	Last, the current prototype becomes easily cluttered when displaying a great amount of labels.
	Several strategies could be used, such as a better clustering of spatially close results, shorter labels, or better label placement.
	
	\subsubsection{Integrating user correction into historical sources}
	In collaborative editing, edits come from untrusted sources.
	Validating edits and solving conflicts is a classic problem.
	In our prototypes, every user edit is potentially used by the geocoder (they are added to a dedicated gazetteer).
	We could use a voting scheme where edits are only taken into account when a sufficient number of users have made the same ones.
	However, we stress that due to the number of data to edit (several hundred thousands of building numbers), we prefer to rely on user benevolence, considering that a user who decides to spend the time to edit century old historical data are committed to accurate editing.
	\subsubsection{Scalability}
	The main design choice of our geocoding architecture is to use a flat model for the address (an address is any set of characters), as opposed to current geocoders which are highly hierarchical (an address refers to a street, which refers to a neighbourhood, etc.).
	This modelling choice allows for the necessary freedom for incomplete historical data, but also comes with a trade-off regarding scaling capabilities.
	Indeed, for strongly hierarchical data, it is possible to have separate databases for each city for instance, thus preventing one database to grow too much, and ensuring a nice scaling capability.
	
	However, this is not the case with our architecture. By using database indexes, we can theoretically guarantee a fast geocoding time for up to few dozen millions geohistorical objects used as sources.
	The main bottleneck in this case is not the temporal aspect (it relies on PostGIS geometry, which enable multiple theoretical solution for scaling), but the textual aspects (\emph{i.e.} the address string itself).
	To scale over dozens of millions of addresses, specific architectures may be used to deal with the text search, for instance distributed database (database sharing), in a way that resembles the current software Elastic Search.
	We stress however that given the current available amount of historical sources, such a scaling problem should not be an issue before a long time.

	\section{Conclusion}
	This article tackles the historical geocoding problem.
	As shown throughout the article, the historical aspects bring major complications to the geocoding problem.
	The main difficulties come from the nature of historical data (uncertainty, fuzzy date, precision, sparseness), which prevents the use of current-address geocoding methods based on strong hierarchical modelling. 
	Instead, we propose a historical geocoding system based on a sound geohistorical object model.
	This model is designed to cover the minimal features, and, thanks to its genericness, modularity, and open source nature, can easily be extended to fit other historical sources.
	Geohistorical objects from several historical sources have been integrated into the database and coherently georeferenced and edited to form gazetteers.
	Geocoding an address at a given time is a matter of finding the best matching geohistorical object in the gazetteers, if any. 
	Our simple, coherent, historical geocoding system has been tested on several real-life datasets collected by historians and can be easily used for other places/times/types of localisations.
	Diverse historical sources covering two centuries for the city of Paris have been integrated into the geocoder.
	The proposed geocoder is able to localise a large percentage of addresses at a fast pace (about 200ms per address).
	Finally, the article describes a prototype of web-based User Interface which demonstrates the interest of collaborative editing of address localisation, and helps historians and other digital humanities researchers use geocoding services.
	
	\vspace{6pt} 
	
	\supplementary{
		All the code and additional documentation is available on the project websites
		\url{http://geohistoricaldata.org} and its associated code repository 
		\url{https://github.com/Geohistoricaldata}.
		The code for the geocoder itself it available here: \url{https://github.com/GeoHistoricalData/historical_geocoding}.
	}
	
	\acknowledgments{
		Thanks to the Belle Epoque project and Angelo Riva and Thierry Géraud, the Institut Louis Bachelot for funding. Thanks to historians who contributed to creating the datasets, especially Benoit Costes for the edit of Alphand map.
		The authors thank the reviewers for their patient, in-depth reviews and suggestions.
		They led to very substantial and beneficial changes.
	}
	

	

	\externalbibliography{yes}
	\bibliography{lite}
	
	
\end{document}